\documentclass[aps,prl,groupedaddress,showpacs]{revtex4-1}

\usepackage{amsmath,amssymb,mathrsfs,graphicx,comment}
\usepackage[dvipsnames,usenames]{color}
\usepackage{dcolumn}
\usepackage{bm}
\usepackage{ulem}
\usepackage{float}
\usepackage{textcomp}

\newcommand{\ket}[1]{| #1 \rangle}

\newcommand{\st}[1]{_\mathrm{#1}}

\newcommand{\Wpcms}{\,W/cm$^2$}

\begin{document}
\title{The response of a neutral atom to a strong laser field probed by transient absorption near the ionisation threshold}

\author{E.R. Simpson}
\author{A. Sanchez-Gonzalez}
\author{D.R. Austin}
\author{Z. Diveki}
\author{S.E.E. Hutchinson}
\author{T. Siegel}
\author{M. Ruberti}
\author{V. Averbukh}
\author{L. Miseikis}
\author{C. Str\"{u}ber}
\author{L. Chipperfield}
\author{J.P. Marangos}
\email{j.marangos@imperial.ac.uk}
\affiliation{Imperial College London, Department of Physics, Faculty of Natural Sciences, Prince Consort Road, London SW7 2BW, UK}

\begin{abstract}

We present transient absorption spectra of an extreme ultraviolet attosecond pulse train in helium dressed by an 800\,nm laser field with intensity ranging from $2\times10^{12}$\,W/cm$^2$ to $2\times10^{14}$\,W/cm$^2$. The energy range probed spans 16--42\,eV, straddling the first ionisation energy of helium (24.59\,eV). By changing the relative polarisation of the dressing field with respect to the attosecond pulse train polarisation we observe a large change in the modulation of the absorption reflecting the vectorial response to the dressing field. With parallel polarized dressing and probing fields, we observe significant modulations with periods of one half and one quarter of the dressing field period. With perpendicularly polarized dressing and probing fields, the modulations of the harmonics above the ionisation threshold are significantly suppressed. A full-dimensionality solution of the single-atom time-dependent Schr\"odinger equation obtained using the recently developed ab-initio time-dependent B-spline ADC method reproduce some of our observations. 

\end{abstract}
\maketitle

\section{Introduction}

Dressing an atom with a laser field changes its response to light in diverse ways. Weak resonant fields drive transitions between the field-free atomic states, enabling resonant atomic coherent phenomena such as electromagnetically induced transparency \cite{Boller-1991-Observation,Marangos2005}, Autler-Townes splitting \cite{Wu-2013-Time,Pfeiffer-2012-Transmission} and related phenomena such as slow light \cite{Hau-1999-Light} to be observed. Stronger dressing fields can modify the optical response without inducing population transfer. By recording the absorption spectrum of an extreme ultraviolet (XUV) attosecond pulse or pulse train in the dressed medium, the response can be studied on timescales below the period of the dressing field. This technique of attosecond transient absorption spectroscopy (ATAS) \cite{GoulielmakisNature2010,Wang-2010-Attosecond,Santra-2011-Theory} has recently been used to study phenomena ranging from electronic coherence in atoms, to ultrafast switching in dielectrics and the lifetime of autoionizing states. In general it holds great promise for monitoring dynamics on a sub- or few-femtosecond timescale in both gas and condensed phase media. 

Despite the simplicity of its structure helium  exhibits rich dynamics and has received much attention. The double excitation around 60\,eV serves as a model system for observing and manipulating a two-electron wavepacket \cite{Ott-2014-Reconstruction}. Even single excitations around the first ionisation threshold present a complex array of effects due to the various bound state resonances. The XUV pulses used in ATAS, produced by high-order harmonic generation (HHG), are typically too weak to induce multi-photon effects by themselves. Therefore, the diverse phenomena reported thus far  can be broadly categorized by the degree of perturbation induced by the dressing field. At the lowest extreme, two photon XUV+IR transitions to dipole forbidden states have been observed at intensities as low as $0.5\times10^{11}$\,W/cm$^2$ \cite{ChenPRL2012,Bell-2013-Intensity,ChiniSciRep2013}. At slightly higher dressing field intensities three photon XUV+2IR transitions occur, enhanced by bound-state resonances, and can undergo ``which-way'' quantum interference with single photon XUV transitions \cite{ChenPRA2013,ChiniSciRep2013,Wang-2013-Subcycle} or optical interference with incident light \cite{Chini-2014-Resonance,Holler-2011-Attosecond}. Additionally the sub-cycle AC Stark shift modulates the frequency of excited states \cite{ChiniPRL2012} and reduces their lifetime. Both of these effects have been detected at $\approx 3\times10^{12}$\,W/cm$^2$. More recently, processes involving four IR photons have been detected \cite{Herrmann-2015-Multiphoton,Chen-2012-Transient}. 

There have been few experimental studies of intensities above $5\times10^{12}$\,W/cm$^2$. Due to the potential for higher order processes, one might expect the physics to be even richer. One possibility is for XUV initiated high-order harmonic generation (XiHHG), in which a continuum electron is ionized by the XUV pulse, and undergoes acceleration in the laser field before recombining with the ion \cite{SchaferPRL2004,GaardePRA2005}. It has also been shown that HHG from an excited atomic state may lead to efficient control over the rescattering efficiency and harmonic polarisation \cite{Averbukh2004}. XiHHG has been reported at $5\times10^{13}$\,W/cm$^2$ \cite{GademannNJP2011}, and has also been proposed as a probe of core dynamics such as Auger decay \cite{LeeuwenburghPRL2012}. 
However, at these intensities, resonant enhancements \cite{Chen-2012-Transient} are expected to be weakened by appreciable broadening of the bound states \cite{Shivaram-2012-Attosecond}. At sufficiently high intensities, a trajectory-based view is likely to become applicable, in which the excited state dynamics are dominated by acceleration in the laser field in a similar fashion to the widely used strong-field approximation \cite{Corkum-1993-Plasma,Krause-1992-High-order,Lewenstein-1994-Theory} for conventional HHG.

Here, we present the first systematic study of the attosecond transient absorption of laser-dressed helium at photon energies around the ionisation threshold and covering a range of laser intensities from $2\times10^{12}$\,W/cm$^2$, in the perturbative regime, to $2\times10^{14}$\,W/cm$^2$, just below the onset of strong-field ionisation. We also present the first anisotropy measurements by comparing the absorption with the XUV and laser fields parallel and perpendicularly polarized. We use an attosecond pulse train, so our measurements are sensitive to the components of the polarisation with the same half-cycle periodicity as the driving laser \cite{Chen-2012-Transient}. We also present full-dimensionality (3-D) time-dependent Schr\"odinger equation calculations performed using the recently developed fully ab-initio time-dependent B-spline ADC method \cite{RubertiJChemPhys2014}, which reproduces some of our observations.

\section{Experimental Methods}

\begin{figure}
	\centering
\includegraphics[trim=0cm 0cm 0.5cm 0.5cm,clip,width=0.70\linewidth]{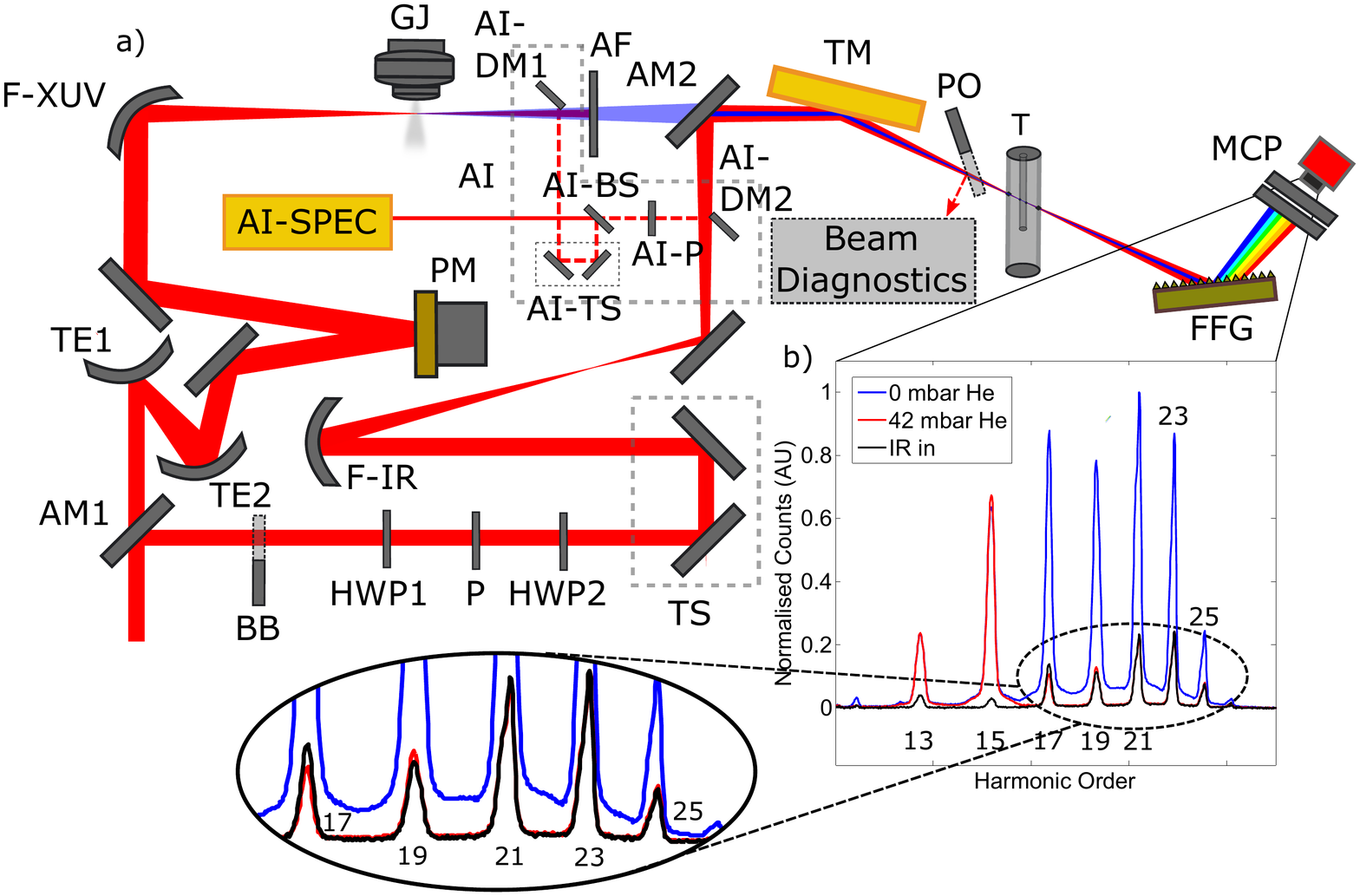}
	 \caption{(a) Schematic diagram of the experimental setup. The incident angle for zero-degree optics was minimised with additional mirrors not shown in this schematic. The parity of the arms was matched to minimise effects from drifts in beam pointing. The label definitions are as follows: annular mirror (AM), insertable beam block (BB), half-wave plate (HWP), polariser (P), piezo-driven mirror (PM), translation stage (TS), focusing optic (F), (TE1) and (TE2) form an expanding telescope, effusive krypton gas jet (GJ), aluminium filter (AF), toroidal mirror (TM), insertable pick-off (PO), differentially pumped helium tube-target (T), flat-field grating (FFG), micro-channel plate (MCP), auxiliary interferometer (AI), D-shaped mirror (DM), beam splitter (BS), spectrometer (SPEC).  (b) Example comparison of the source harmonics (blue) generated in the effusive gas jet (GJ) with no gas in the target (T) compared with the field-free absorption (red) of the source harmonics in the undressed helium target (T).  The harmonic transmission of the laser dressed helium is also shown (black) for a given delay and peak intensity $1.2\times10^{14}$\,W/cm$^2$. The above ionisation energy harmonics (17--25) are shown in the zoomed insert.}
	\label{fig:setup}
\end{figure}

Figure~\ref{fig:setup} shows a schematic for our experimental setup for measuring transient absorption. The input pulses, of wavelength 800\,nm, full-width at half maximum duration (FWHM) 30\,fs and energy $\sim$3\,mJ, were supplied by a 1kHz, Ti:Sapphire CPA laser (KMLabs Red Dragon). An annular mirror (AM1)  split the incoming beam. The transmitted portion was used to produce the XUV attosecond pulse train using HHG in an effusive gas jet (GJ), with a 100\,$\mu$m diameter nozzle backed by 1.5\,bar krypton. The residual IR was blocked by an aluminium filter (AF). The beam reflected from (AM1) was used to dress the helium atoms. Its intensity was controlled using a half-wave plate (HWP1) and polarizer (P), and its polarisation was switched between vertical and horizontal by a further half-wave plate (HWP2). An insertable beam block (BB) was used to block the dressing field beam for reference (field-free absorption) spectra. The time delay between the XUV and dressing arms was controlled with either a delay translation stage (TS) or piezo-driven mirror (PM). An auxiliary interferometer (AI) provided high-resolution tagging for the delay between the two arms.   The arms were recombined using a further annular mirror (AM2) and refocused with a toroidal mirror (TM) into a 2.6\,mm diameter helium-filled tube target (T).  The transmitted XUV spectrum was dispersed by a 1200\,lines/mm flat-field grating (FFG) and detected on a micro-channel plate (MCP) with phosphor coating, monitored by a CCD camera.  Around the ionisation threshold of helium, the individual harmonics were well resolved. The measured spectral range was 16--42\,eV, corresponding to harmonic orders 11--27.

The peak intensity of the dressing beam in the target plane was inferred from power and beam profile measurements, onto which the beam was directed by an insertable pick-off (PO). The FWHM of the dressing beam focus was $\approx$ 100\,\textmu m from focal spot imaging. The spot size of the XUV beam was $\approx$ 70\,\textmu m FWHM, measured using a knife edge. We estimate that there is a large uncertainty in the absolute value of the intensity ($\pm$\,50\,$\%$), however relative intensity measurements are determined with much less uncertainty. The experimental pulse duration of the dressing field was estimated to be $\sim$55\,fs FWHM from cross-correlation measurements, and increasing positive delay corresponds to a later arrival of the XUV pulse.

The temporal resolution was taken as the root-mean-square fluctuation of the delay between the XUV and dressing arms as measured in the auxiliary interferometer. Over the duration of one exposure of the MCP camera this was 150\,as placing an upper limit on the observable modulation frequency of $\approx 8\,\omega_1$, where $\omega_1=2.35$\,rad/fs is the dressing field frequency.

To convert our delay-dependent spectra into absolute absorption cross sections, we first determined the density-length product $\eta$ of the target, $150$\,mbar$\cdot$mm equivalent to $0.37$\,Mb$^{-1}$, from the field-free absorption and its known absolute cross section $\sigma\st{FF}$ \cite{Baker1961}. 
The delay-dependent cross section was then given by
\begin{equation}
\sigma(\tau)=\eta^{-1}\log{\frac{I\st{FF}}{I(\tau)}}+\sigma\st{FF}
\end{equation}
where $I\st{FF}$ and $I(\tau)$ are the measured field-free and delay-dependent intensities.

\section{Theoretical Methods}

We calculated the  single-atom delay-dependent absorption using   
 B-spline time-dependent (TD) algebraic-diagrammatic construction (ADC) method \cite{RubertiJChemPhys2014}. The basis set consists of spherical harmonics for the angular part and B-splines $B_i(r)$
for the radial coordinate. The single particle basis functions $\psi_{ilm}$ used in this calculation are therefore expressed as
\begin{equation}\label{bsplines}
\psi_{ilm}=\frac{1}{r} B_{i}(r) Y_{lm}(\theta,\phi).
\end{equation}

The time-dependent problem is solved within TD-ADC making the following ansatz for the time-dependent many-electron wavefunction $\ket{\Psi (t)} $
\begin{equation}\label{CESADC}
\ket{\Psi(t)} = C_{0}(t) \ket{ \Psi_{0}(t)}  + \underset{n}{\sum} C_{n}(t) \ket{ \Psi_{n}(t)},
\end{equation}

where the coefficients $C_{0}(t)$ and $C_{n}(t)$ refer to the ground-state $\ket{\Psi_{0}}$ and to the correlated excited states (CES) $\ket{\Psi_{n}}$  of 
the ADC theory respectively. These configuration basis states include single, double, etc. excitations with respect to the ground state of the system; the maximum number of electrons which are
allowed to be excited at the same time, i.e. the point at which the expansion of Eq.~\ref{CESADC}  is truncated, defines the order n of the ADC(n) hierarchy.
In the following calculation we have used the first-order method of the ADC-hierarchy, namely ADC(1), in which only single excitations are included in the expansion of
the wavefunction. This is a good approximation to the current application as in He the threshold for double excitation is above 60\,eV and the photon energies of the harmonics investigated here are below 40\,eV. The presented results have been calculated making explicit use of the atomic spherical symmetry and they are
in principle exact as long as double excitations do not play an important role in the dynamical process of interest.
The time-dependent Schr\"odinger equation (TDSE) for the unknown coefficients $C_{0},C_{n}$ is solved via
the Arnoldi-Lanczos algorithm. A complex absorbing potential (CAP) has been employed in order to eliminate wave-packet reflection effects from the grid boundaries.

The frequency dependent absorption is calculated from the expectation
value of the electric dipole moment of the atom $z(t)$ and the incident XUV field $E\st{X}(t)$ as

\begin{equation}
S(\omega;\tau_{d}) = - \mathrm{Im} \left[\tilde{E}\st{X}^{\star} (\omega)  \langle \tilde{z} (\omega;\tau_{d})\rangle \right],
\end{equation}

where tilde denotes the Fourier transform from time to XUV frequency $\omega$, star denotes complex conjugation, and $\tau\st{d}$ is the IR-XUV delay. This quantity is then Fourier transformed with respect to  $\tau\st{d}$ to even multiples of the IR frequency $\omega_1$. 

The generalised cross-section, $\sigma(\omega;\tau_{d})$ was calculated using the following equation \cite{Gaarde2011}

\begin{equation}
\sigma(\omega;\tau_{d}) = \frac{4\pi\alpha\omega S(\omega;\tau_{d})}{|\tilde{E}\st{X}|^{2}},
\end{equation}
where $\alpha$ is the fine structure constant.

\section{Results and Discussion}

\begin{figure}
\includegraphics[trim=0cm 1cm 0cm 1cm]{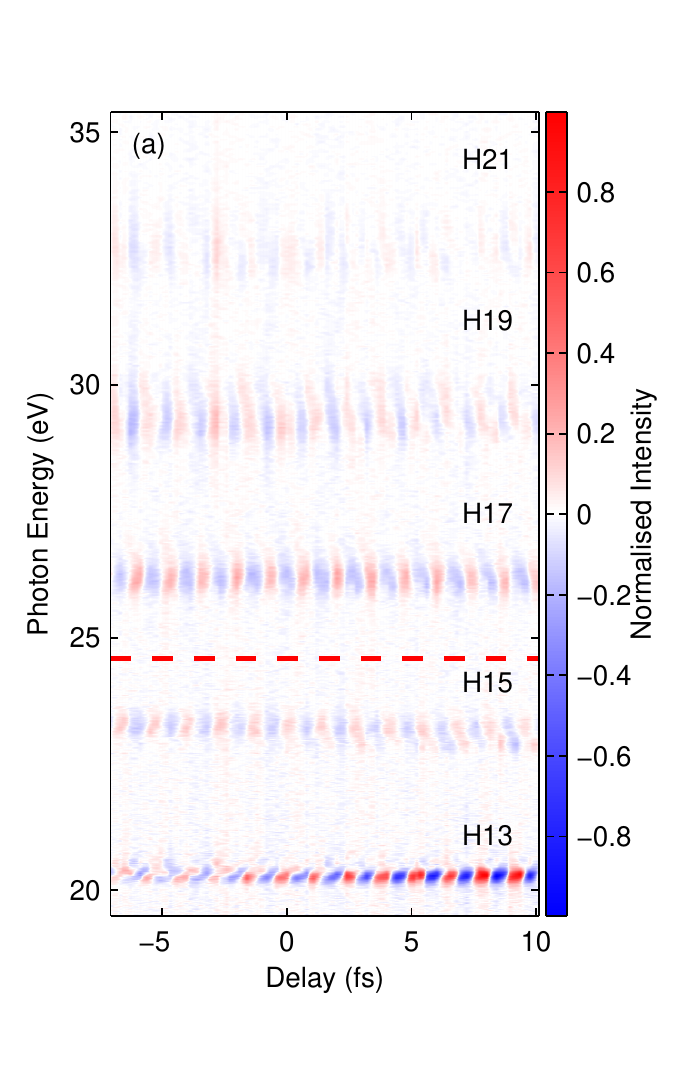} 
	\hspace{2cm}
	\includegraphics[trim=0cm 1cm 0cm 1cm]{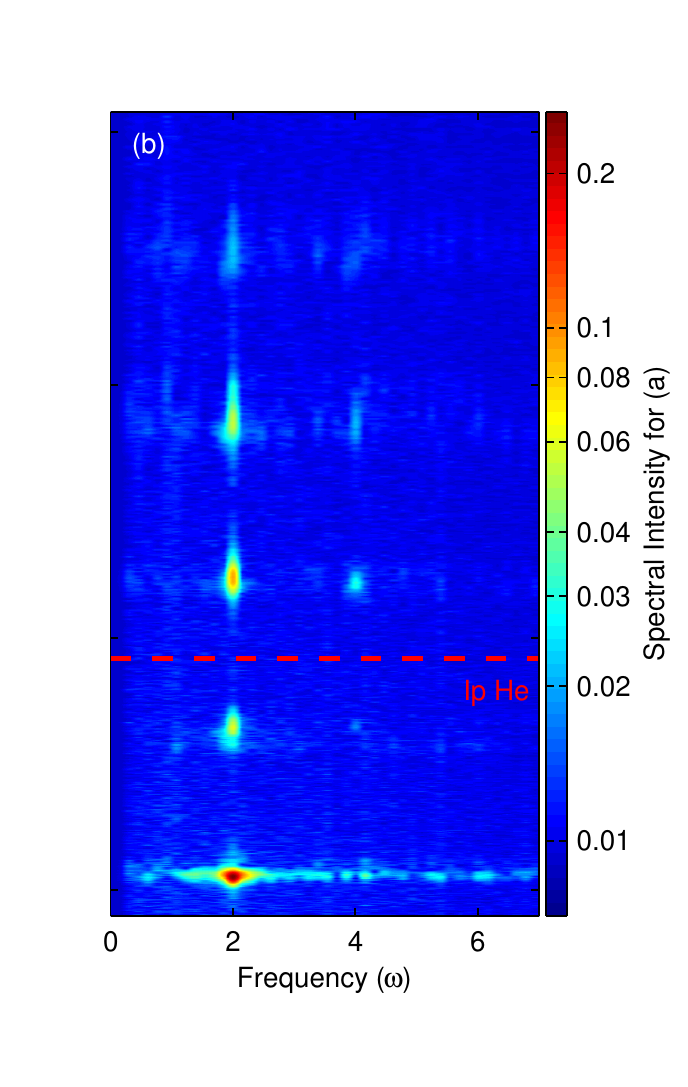} 
	\caption{(a) Transmitted XUV pulse train spectrum with dressing field intensity $2.2 \times 10^{14}$\,W/cm$^2$. High-pass filtering along the time-delay axis has been applied to remove background drift and isolate oscillating components. (b) Fourier transform along the time-delay axis of (a). The dashed line indicates the first ionisation potential of helium. }
	\label{fig:fig2}
\end{figure}

Figure~\ref{fig:fig2}(a) shows the results of one delay scan at a dressing field intensity of $2.2\times10^{14}$\,W/cm$^2$ ($\sim$3.8\,mJ input pulses) with parallel polarized XUV and dressing fields. Clear modulation is observed for harmonics 13--21, (20--32.5\,eV), spanning the ionisation potential of helium. The Fourier transform along the delay axis, shown in figure~\ref{fig:fig2}(b), reveals that while the modulation is predominately at twice the dressing field frequency, harmonics 17 and 19 also have significant modulation at $4\omega_1$.  Since the modulation frequency is determined by the difference in the frequencies of the harmonics coupled by the dressing field \cite{Chen-2012-Transient}, we infer that there is strong coupling between all adjacent harmonics. Given H11 was blocked by the aluminium filter, the entire $2\omega_1$ component of H13 must be due to coupling with H15, whereas the $2\omega_1$ components of the other harmonics are coherent combinations of the couplings between both adjacent harmonics. The $4\omega_1$ component implies weaker but significant coupling between H17 and H13 or H21 (or both) and also between H19 and H15 or H23 (or both).

\begin{figure}
	\centering
\includegraphics[trim=0cm 0.2cm 0cm 1cm]{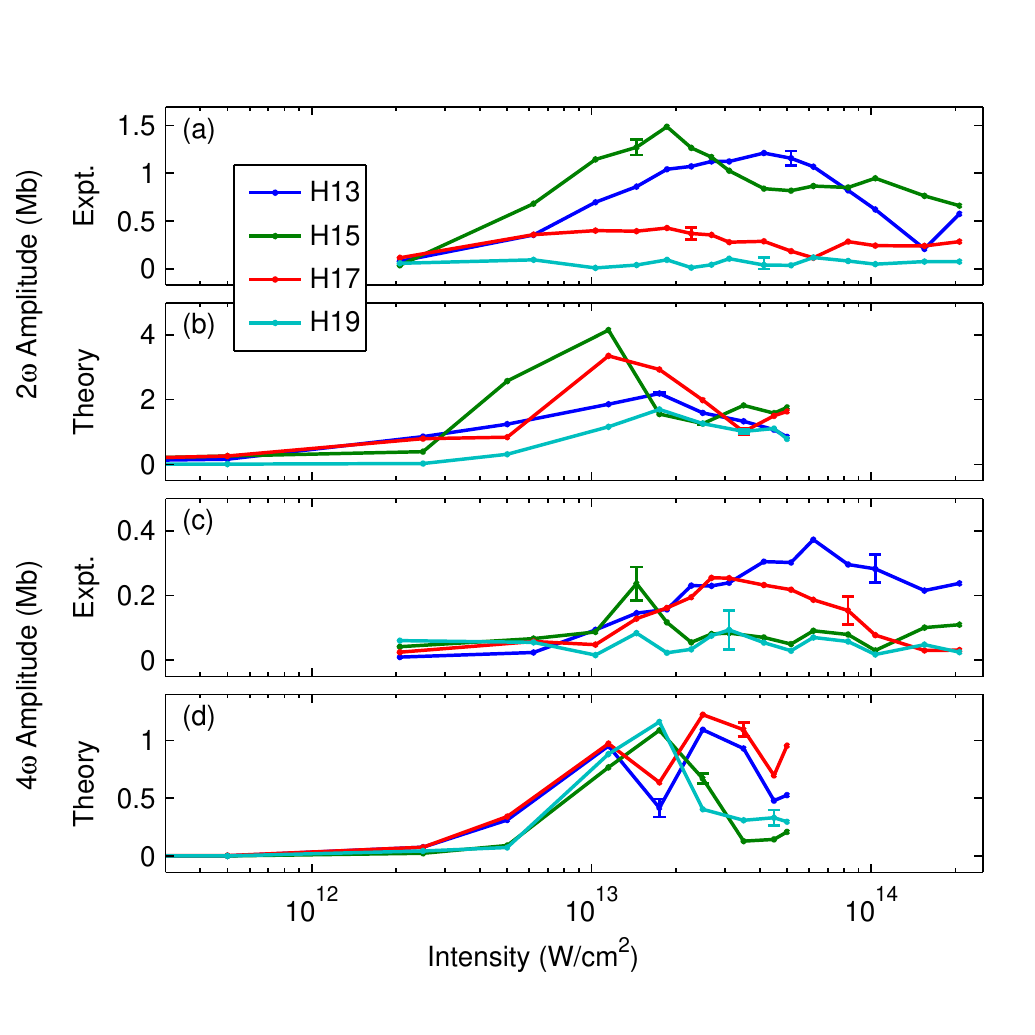}
		\caption{Dependence of absorption cross section modulation amplitudes on intensity; (a) and (c) show experimental results, (b) and (d) show theory. (a), (b): Half-cycle period component. (c), (d): Quarter-cycle period component. Representative error bars are shown.} 
		\label{fig:intscan}
\end{figure}

Further insight is obtained from the dependence of the modulation amplitudes on the dressing field intensity. Figure~\ref{fig:intscan}(a) and (c) show the measured amplitude of the $2\omega_1$ and $4\omega_1$ components respectively for harmonics 13--19.
 Across the full intensity range of $2.1\times 10^{12}$\,W/cm$^2$--$2.1\times 10^{14}$\,W/cm$^2$ we detected $2\omega_1$ modulations, with the maximum amplitude occurring around $2\times 10^{13}$\,W/cm$^2$. We also detected $4\omega_1$ modulations above $10^{13}$\,W/cm$^2$ in harmonics 13 and 17.
The results of the corresponding theoretical calculations are shown in figure ~\ref{fig:intscan}(b) and (d). 
The theory reproduces several aspects of the experiment: the existence and approximate 
intensity of the maximum $2\omega_1$ and $4\omega_1$ modulation amplitudes for H15, as well as the higher intensity onset of the $4\omega_1$ modulations. In general, the theoretical amplitudes are much larger than those observed. Possible reasons for the discrepancy include the spatial variation of the dressing field intensity over the XUV beam, and the propagation in the target medium which can reduce or eliminate modulations \cite{Chen-2012-Transient}. For this figure, the theoretical cross sections were obtained following the same procedure as for the experimental data. The theoretical FWHM duration of the IR dressing field pulse was taken as 50\,fs throughout.

\begin{figure}
\includegraphics[trim=0cm 0cm 3cm 0.5cm]{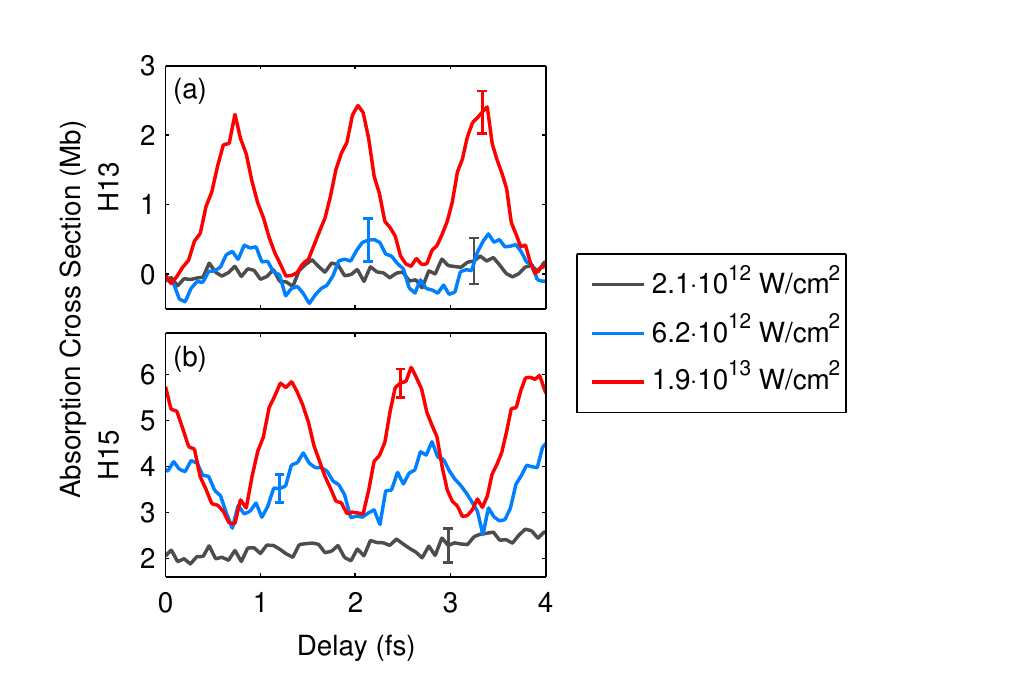}
\caption{\label{fig:H13H15}Delay-dependent absorption of (a) harmonic 13 and (b) harmonic 15 at $2.1\times 10^{12}$
\Wpcms{} (dark grey), $6.2\times 10^{12}$\Wpcms{} (blue) and $1.9\times 10^{13}$\Wpcms{} (red). Representative error bars are shown.}
\end{figure}

We noticed that at certain time delays and for a small range of intensities, the absorption of harmonic 13 was negative, meaning that transmitted light was stronger than if the target were absent. This is examined in greater detail in figure~\ref{fig:H13H15}. Note the relative phases of modulations between different intensity data sets cannot be reliably established in these measurements, but the relative phases of harmonics at each intensity is well defined. We focus on harmonics 13 and its neighbour 15 (harmonic 11 was blocked by the aluminium filter). At $2.1\times10^{12}$\,W/cm$^2$ (dark gray), the absorption is negligible as neither of the harmonics overlap with any bound states. Increasing the intensity to $6.2\times10^{12}$\,W/cm$^2$ introduces a two-IR-photon coupling between these harmonics, causing the absorption of H13 to oscillate around a mean of zero. This is also seen in harmonic 15, but because it is closer to the ionisation threshold its absorption also acquires a cycle-averaged offset of 3.7\,Mb. Increasing the intensity further ($1.9\times10^{13}$\,W/cm$^2$) strengthens the coupling and hence the modulation amplitude, but also opens up an absorption channel for harmonic 13 which offsets the coupling so that absorption becomes positive for all delays.

\begin{figure}
	\centering
\includegraphics[trim=1cm 17cm 0cm 2cm]{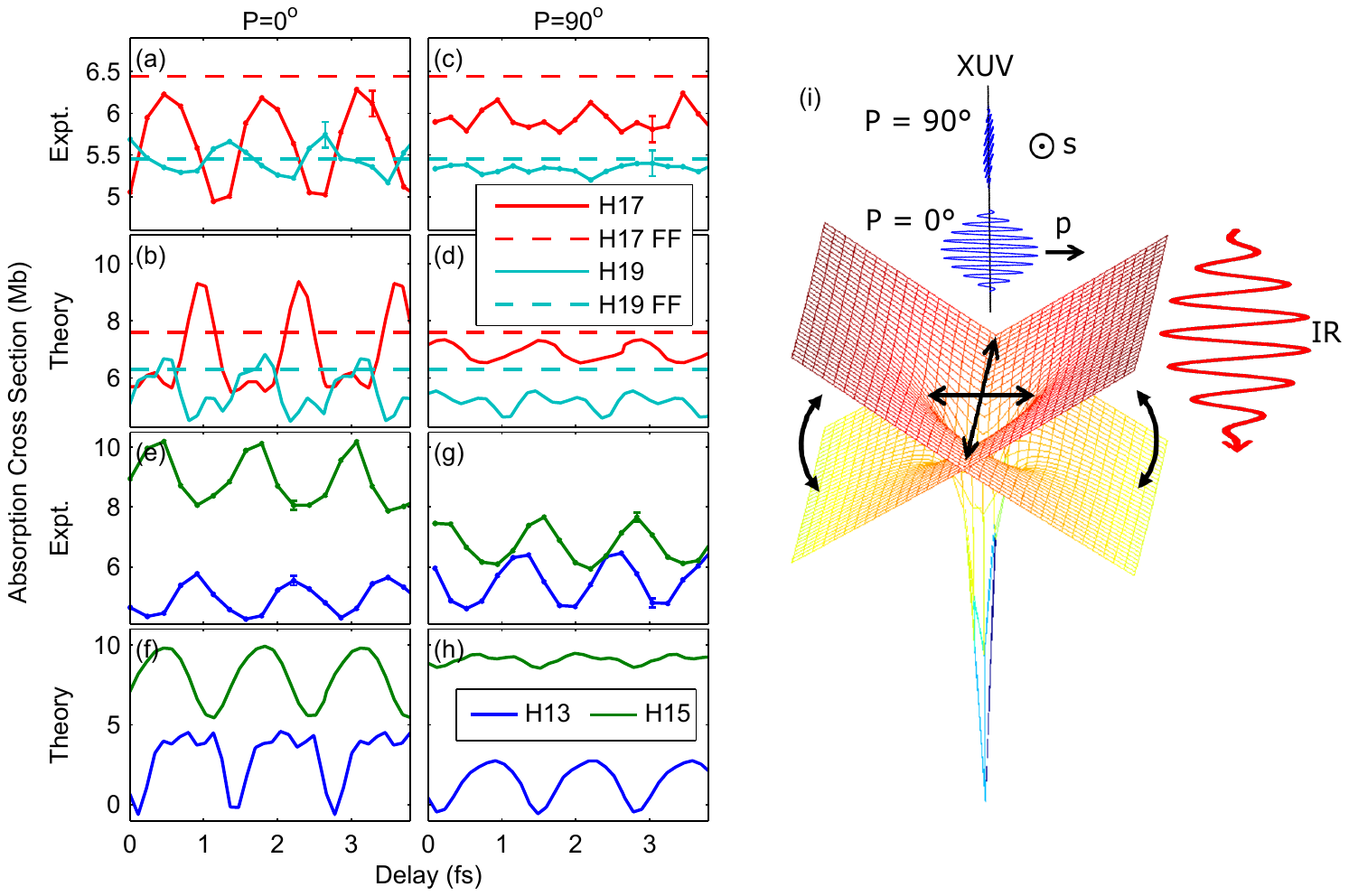}
\caption{\label{fig:polscan}Delay-dependent absorption for parallel (left column) and perpendicularly (right column) polarized dressing and probe fields, for (a)--(d) above-threshold harmonics 17  (red) and 19 (cyan), and (e)--(h) below-threshold harmonics 13 (blue) and 15 (green). Representative error bars are shown. The dashed lines show the experimental and theoretically calculated field-free (FF) absorption. (i) Cartoon picture of the relative polarisation sensitivity of the modulation amplitudes.}
	\label{fig:fig4}
\end{figure}

Figure~\ref{fig:polscan} presents the results for the polarisation dependence with dressing field intensity $5\times10^{13}$\,W/cm$^2$ for the theoretical calculation, and $1.2\times10^{14}$\,W/cm$^2$ for the experiment. Changing the relative polarisation of the dressing and XUV fields from parallel (left column) to perpendicular (right column) causes a significant reduction in the modulation amplitude of the above ionisation energy harmonics 17 and 19 (figure~\ref{fig:fig4}(a) and (c)). This reduction is reproduced by the numerical model (figure~\ref{fig:fig4}(b) and (d)). By contrast, the modulation amplitude of the below threshold harmonics is not reduced by changing the polarisation (Fig.~\ref{fig:fig4}(e) and (g)). This behaviour is partially reproduced by the model (figure~\ref{fig:fig4}(f) and (h)); the modulation amplitude for harmonic 13 is reproduced well however that for harmonic 15, which lies closest to the ionisation energy of helium, is not well captured. The decreased modulation amplitude of the above threshold harmonics as the relative polarisation is changed from parallel to perpendicular can be understood by considering the direction of the dipole induced by the XUV (figure~\ref{fig:fig4}(i)). If the two fields are parallel, the XUV-induced dipole will experience maximum modulation from the IR-induced distortion to the bound potential. When the two fields are aligned perpendicularly, the XUV-induced dipole will experience minimum modulation of the bound potential resulting in a reduction in the absorption modulation amplitude. Alternatively, these observations may be partially explained by a simple trajectory picture invoking the SFA limit, in which the excited (non-ground) part of the wavefunction is viewed as a particle created at the origin with momentum imparted by the XUV field. With parallel polarisation, both the initial momentum and the acceleration from the laser field are in the same direction, and at certain time delays overlap with the ground state and hence recombination can occur. For perpendicular polarized fields, the momentum along the XUV polarisation axis is only affected by the atomic potential. If the initial kinetic energy is above the ionisation threshold, recombination is impossible so the absorption is not strongly modulated. In contrast, for initial kinetic energies below the ionisation threshold, the interplay of the initial momentum, the acceleration in the laser field, and the potential may allow recombination. The varying agreement for the below threshold numerical results may be attributed to the dimension of the radial box used to define the B-spline basis set. Although the finite box can well describe the bound states of helium, diffuse Rydberg states are less well captured. Note the relative phase between theory and experiment cannot be directly compared.

A question arising from our results is the absence of a clear signature of XiHHG. Although this process is understood using SFA picture/trajectories, it can also be recast in a multi-photon picture. Provided the harmonic emitted by the recombining electron is present in the incident field, the emission and incident light should interfere producing delay-dependent absorption with modulation frequency determined by the frequency difference between the initiating and emitting harmonics. Therefore, our observation of $4\omega_1$ modulations at H19 is consistent with HHG initiated by harmonic 15 with a laser-driven kinetic energy gain of $4\omega_1$.
However, our wavelength and typical laser intensity were such that the available kinetic energy gain using the standard recollision model was $3.2 U\st{p} \approx 12\omega_1$, and our best estimates of the temporal resolution yielded a maximum resolvable modulation frequency of $8\omega_1$. An XiHHG picture would therefore predict higher order modulations at the above threshold harmonics, which we did not observe. One possible explanation for the absence of this, as well as other disagreements with theory,  is reshaping of the pulse in the medium. At maximum IR intensity, the absorption of the harmonics was $\sim$80\% or more. Such a degree of absorption is accompanied by significant phase slip between the dressing field IR and the harmonics, and dispersion of the harmonics. This has been shown to be capable of modifying or even removing certain modulation components 
\cite{Chen-2012-Transient} 
and changing absorption line shapes of bound state resonances
\cite{Pfeiffer-2013-Alternating}. Physically, different dynamics are initiated and probed throughout the target length.  Another source of disagreement was the spatial intensity variation of the dressing field in the transverse plane. This is caused by the XUV focal spot size being a significant fraction of the dressing field focal spot size, as well as uncertainty in the spatial and temporal overlap between the XUV and IR beams.

\section{Conclusion}
In summary, we have presented the first measurements of the transient absorption of singly-excited helium dressed by a laser field with intensity approaching the strong-field ionisation threshold of the ground state. We observed delay-dependent absorption with quarter-cycle periodicity and a strong anisotropic dependence on the relative polarisations of the dressing and probing fields. Several aspects of our results were reproduced by single-atom time-dependent Schr\"odinger equation calculations.  We have studied the dependence of field induced absorption modulation on the relative polarisation of dressing and probe fields. This shows a significant modulation for the case of parallel polarisation, but is strongly suppressed in the case of perpendicular polarisations for the above threshold harmonics. This behaviour is reproduced in the calculation. Little effect on the modulation amplitude was observed between the parallel and perpendicular polarisations for the below threshold harmonics.  This behaviour is reproduced in the numerics for the lower energy harmonic but not well captured for the harmonic energy closest to threshold. We understand this to be due to a limited radial box size for the B-spline basis set used in the calculation.

\section{Ackowledgements}

We acknowledge the support from Engineering and Physical Sciences Research Council (UK) (EPSRC) grant EP/I032517/1 and the European Research Council (ERC) ASTEX project 290467. A. S-G. is funded by Science and Technology Facilities Council (STFC).

\section{References}

\bibliography{xihhg}

\end{document}